\documentclass[10pt,twocolumn,letterpaper]{article}

\usepackage{iccv}
\usepackage{times}
\usepackage{epsfig}
\usepackage{graphicx}
\usepackage{amsmath}
\usepackage{amssymb}
\usepackage{authblk}

\usepackage[breaklinks=true,bookmarks=false]{hyperref}

\iccvfinalcopy 

\def\iccvPaperID{4} 

\ificcvfinal\fi

\begin{document}

\title{Domain-Agnostic Learning with Anatomy-Consistent Embedding for Cross-Modality Liver Segmentation}
\author[1]{ Junlin Yang}
\author[3]{ Nicha C. Dvornek}
\author[1]{ Fan Zhang}
\author[1]{ Juntang Zhuang}
\author[3]{Julius Chapiro}
\author[3]{MingDe Lin}
\author[1,2,3,4]{James S. Duncan}
\affil[1]{ Department of Biomedical Engineering, Yale University}
\affil[2]{ Department of Electrical Engineering, Yale University}
\affil[3]{ Department of Radiology \& Biomedical Imaging, Yale School of Medicine}
\affil[4]{ Department of Statistics \& Data Science, Yale University}

\maketitle
\ificcvfinal\thispagestyle{empty}\fi

\begin{abstract}

Domain Adaptation (DA) has the potential to greatly help the generalization of deep learning models. However, the current literature usually assumes to transfer the knowledge from the source domain to a specific known target domain. Domain Agnostic Learning (DAL) proposes a new task of transferring knowledge from the source domain to data from multiple heterogeneous target domains. In this work, we propose the Domain-Agnostic Learning framework with Anatomy-Consistent Embedding (DALACE) that works on both domain-transfer and task-transfer to learn a disentangled representation, aiming to not only be invariant to different modalities but also preserve anatomical structures for the DA and DAL tasks in cross-modality liver segmentation. We validated and compared our model with state-of-the-art methods, including CycleGAN, Task Driven Generative Adversarial Network (TD-GAN), and Domain Adaptation via Disentangled Representations (DADR). For the DA task, our DALACE model outperformed CycleGAN, TD-GAN, and DADR with DSC of 0.847 compared to 0.721, 0.793 and 0.806. For the DAL task, our model improved the performance with DSC of 0.794 from 0.522, 0.719 and 0.742 by CycleGAN, TD-GAN, and DADR. Further, we visualized the success of disentanglement, which added human interpretability of the learned meaningful representations. Through ablation analysis, we specifically showed the concrete benefits of disentanglement for downstream tasks and the role of supervision for better disentangled representation with segmentation consistency to be invariant to domains with the proposed Domain-Agnostic Module (DAM) and to preserve anatomical information with the proposed Anatomy-Preserving Module (APM).

\end{abstract}

\section{Introduction}
Domain Adaptation (DA) has emerged as an effective technique to help the generalization of deep learning models \cite{wang2018deep}. Although supervised deep learning models have been very successful in a variety of computer vision tasks, such as image classification and semantic segmentation, it usually requires lots of labeled data and assumes that training and testing data are sampled \textit{i.i.d} from the same distribution. In practice, it is expensive and time-consuming to collect annotated data for every new task and new domain. At the same time, domain shift is common, which means training and testing data are typically from different distributions but related domains. 

In medical imaging, domain shift can be caused by different scanners, sites, protocols and modalities, adding to the high cost and difficulties of collecting large medical imaging datasets annotated by experts. Progress has been achieved to tackle this problem, especially for the domain shift caused by different scanners, sites and protocols. Yet, DA between different modalities is more challenging and yet to be extensively explored due to the large domain shift between different modalities \cite{dou2018pnp}. Once achieved, it will not only solve the scarcity of annotated data for medical imaging, but also greatly improve the current clinical workflows and the integration of different modalities. For example, both CT and MR play an important role in the diagnosis and follow-up after treatment of hepatocellular carcinoma (HCC) and they provide entirely different information. MR provides better specificity and multi-parametric tissue characterization along with better soft tissue contrast which helps identify fat, diffusion, and enhancement in a much more dynamic way, while CT merely measures perfusion and density of tissue. CT is quantitative due to calibration of density with Hounsfield unit, while MRI is not \cite{oliva2004liver}. It is desired to achieve domain adaptation from CT to MR since CT is cheaper and more available in practice and many tasks such as liver segmentation are usually required on each modality. 

Most current works on the domain shift problem assume that the target domain is specific and known as a prior and try to adapt the source domain into a distinct target domain. Domain Agnostic Learning (DAL) \cite{peng2019domain} proposes a novel task to transfer knowledge from a labeled source domain to unlabeled data from arbitrary target domains, a difficult yet practical problem. For example, target data could consist of images from different medical sites, from different scanners and protocols, or even from different modalities. The main challenge is that the target data is highly heterogeneous and from mixed domains. 

Mainstream DA methods for semantic segmentation in medical imaging such as CycleGAN \cite{zhu2017unpaired} and its variant TD-GAN \cite{zhang2018task} work at the pixel level. However, they assume a one-to-one mapping between source and target, and thus are unable to recover the complex cross-domain relations in the DAL task \cite{almahairi2018augmented, huang2018multimodal}. Furthermore, the translation in pixel-level information by making the marginal distributions of the two domains as similar as possible does not necessarily guarantee semantic consistency \cite{tzeng2015simultaneous}. This is also the case for methods that incorporate feature-level marginal distributions alignment which do not explicitly enforce semantic-consistency, such as DADR \cite{yang2019domain}.

In this work, we propose an end-to-end trainable model that solves not only the problem of unsupervised DA, but also works for DAL. Our DALACE model learns domain-agnostic anatomical embeddings by disentanglement under the supervision of a Domain Agnostic Module (DAM) and an Anatomy Preserving Module (APM). It enforces semantic-consistency to ensure the disentangled domain-agnostic feature space to be meaningful and interpretable, instead of simply aligning marginal distributions via adversarial training. Our model outperforms the state-of-the-art models on DA and generalizes naturally to the DAL task. We show the success of disentangling anatomical information and modality information by visualization of domain-agnostic images and modality-transferred images. Our model thus improves the interpretability of black-box deep neural network models. Through ablation studies, we show that the performance of the downstream task benefits from the learned disentangled representations, and the proposed supervision modules DAM and APM boost the disentanglement. Furthermore, domain-agnostic images generated by our DALACE model have the potential for training a better joint learning model that utilizes the annotations from all modalities and works the best on each modality at the same time. This initial effort to help the integration of different modalities is valuable, as each modality has its unique strengths and plays its unique role in clinical practice. The main contributions are summarized below.



First, this work explicitly proposes and tackles the DAL task for medical image segmentation. With the supervision of DAM and APM, the proposed end-to-end model learns a domain-agnostic anatomical embedding to reduce the domain shift while preserving the anatomy. 
Second, numerous experiments were conducted to show the effectiveness of our proposed model for the DA, DAL and joint learning tasks with large CT and small MR datasets.  
Third, We show the designed model by disentanglement to be more interpretable through visualization. Ablation studies show the benefit of disentanglement for the downstream task and the role of supervision for disentanglement.

\section{Related Work}
\textbf{Domain Adaptation} has been a popular topic and is the potential solution for generalization of deep learning models. There are mainly two categories, feature-level domain adaptation that aligns features between domains and pixel-level domain adaptation that performs style-transfer between domains \cite{wang2018deep}. For medical images, domain adaptation between different domains caused by different scanners, medical sites and modalities is quite important, considering the high cost of collecting and annotating medical images from different domains and the valuable and unique roles of different modalities in clinical practice. Most state-of-the-art domain adaptation methods for medical image segmentation reduce the domain shift through adversarial learning. For example, CycleGAN \cite{zhu2017unpaired} and its variants TD-GAN \cite{zhang2018task} and TA-ADA \cite{jiang2018tumor} rely on the cycle-consistency loss and have led to impressive results. However, they assume a one-to-one mapping, instead of many-to-many, between data with complex cross-domain relations. Thus, they fail to capture the true structured conditional distribution. Instead, these models learn an arbitrary one-to-one mapping and generate translated output lacking in diversity \cite{almahairi2018augmented, huang2018multimodal}. DADR \cite{yang2019domain} achieves DA by disentangling medical images into content space and style space. However, anatomy-consistency is not always guaranteed without explicitly enforcing semantic consistency on content space. As for feature-level adaptation, while it seems effective for tasks like classification, it is unclear how well it might scale to dense structured domain adaptation \cite{ramirez2018exploiting, luo2019taking}. 

\textbf{Domain Agnostic Learning.} Compared to Domain Adaptation, Domain Agnostic Learning aims to learn from a source domain and map to arbitrary target domains instead of one specific known target domain \cite{peng2019domain}. In the field of medical imaging, it is an interesting task to explore since it is common to get test data from different domains caused by different scanners, sites, protocols and modalities \cite{dou2018pnp}. As for cross-modality liver segmentation, the DAL task is in particular useful since images from many different modalities (e.g. CT, MR with different phases, etc.) are routinely acquired for better diagnosis, image guidance during treatment and follow-up after treatment \cite{oliva2004liver}. Mainstream DA methods align the source and target domains by adversarial training \cite{long2015learning, tzeng2017adversarial}. However, with highly entangled representations, these models have limited capacity to tackle the DAL task. \cite{peng2019domain} proposes to solve the DAL task for classification by learning disentangled representations. 


\textbf{Disentangled Representation Learning.} Disentangled representation learning aims to model the different factors of data variation \cite{higgins2018towards}. A couple of methods have been proposed to learn disentangled representations \cite{chen2016infogan, higgins2017beta}. Some focus on disentangling style from content \cite{huang2018multimodal}. In our case, we define content as anatomy information, i.e., spatial structure, and define style as modality information, i.e., the rendering of the image. Recent work \cite{locatello2019challenging} suggests that future research on disentangled representation learning should investigate concrete benefits of enforcing disentanglement of the learned representations and be explicit about the role of inductive biases and supervision. In our work, we discuss the performance boost by disentanglement learning and the role of supervision from our proposed anatomy preserving module (APM) and domain agnostic module (DAM) through ablation studies. Disentanglement learning also plays an important role to go from the DA task to DAL task. 


\textbf{Interpretation by Disentanglement} Deep neural networks are generally considered black box models. However, there has been lots of recent work on interpretation of deep learning models, particularly in medical imaging \cite{li2019efficient, codella2018collaborative}. \cite{gilpin2018explaining} summarizes these works into three main categories, including emulating the processing of the data to draw connections between the inputs and outputs, explaining the representation of data inside the network, and designing neural networks to be easier to explain. Disentangled representation falls into the last sub-category since these networks are designed to explicitly learn meaningful disentangled representations \cite{chen2016infogan, higgins2017beta}. Through visualization of transferred images and domain-agnostic images and experiments on downstream tasks, we not only show the success of disentanglement between anatomy information and modality information, but also show the representation has the potential for task transfer and data reconstruction. Furthermore, experimental results demonstrate that the downstream tasks benefit from the learned disentangled representation. These results show that our model is designed to learn meaningful, interpretable representations.

\section{Method}
\begin{figure}[t]
\begin{center}
\includegraphics[width=0.623\linewidth]{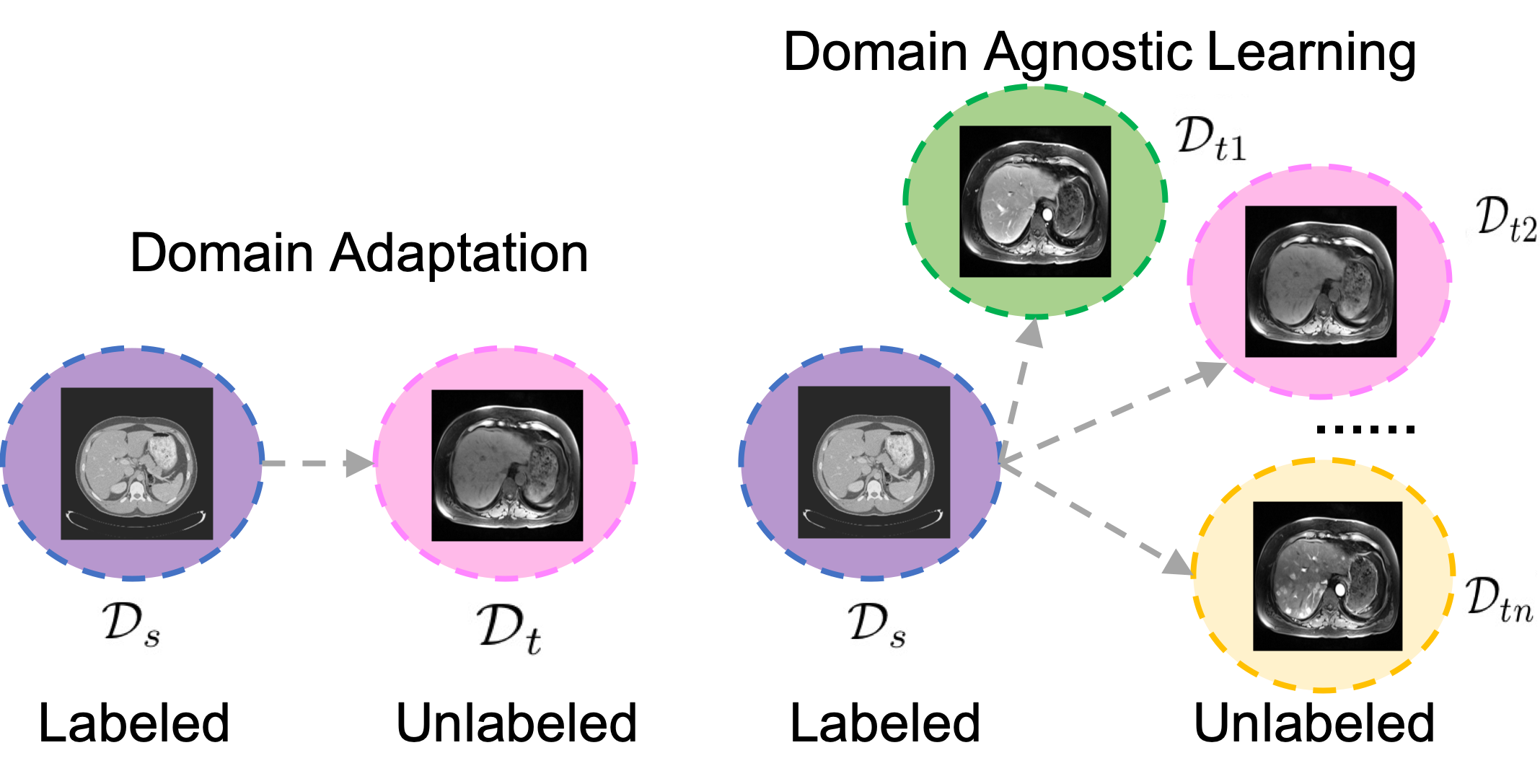}
\end{center}
   \caption{Schematic diagram of the domain adaptation task and the domain agnostic learning task.}
\label{fig:DA_DAL}
\end{figure}

We propose an end-to-end trainable Domain Agnostic Anatomical Embedding by Disentanglement (DALACE) model to tackle the DA and DAL tasks. Of note, the DA task is defined as transferring knowledge from a given source labeled dataset belonging to domain $\mathcal{D}_s$ to a target unlabeled dataset that belongs to a specific known domain $\mathcal{D}_t$. The DAL task is defined in a similar way, except that the target unlabeled dataset consists of data from multiple domains $\{\mathcal{D}_{t1}, \mathcal{D}_{t2}, ..., \mathcal{D}_{tn}\}$ without any domain label for each sample annotating which domain it belongs to. The ultimate goal is to minimize the target risk for downstream tasks \cite{peng2019domain}. In our application to medical images from patients with HCC, CT has true segmentation masks while MR does not. CT and MR are unpaired with each other. The DA task is to transfer knowledge from CT data to pre-contrast phase MR data, the DAL task is to transfer knowledge from CT data to heterogeneous multi-phasic MR data from mixed domains, and the downstream task of interest is cross-modality liver segmentation. Please see the visualization of DA and DAL in Fig. \ref{fig:DA_DAL}.

\begin{figure*}[t]
\begin{center}
\includegraphics[width=0.78\linewidth]{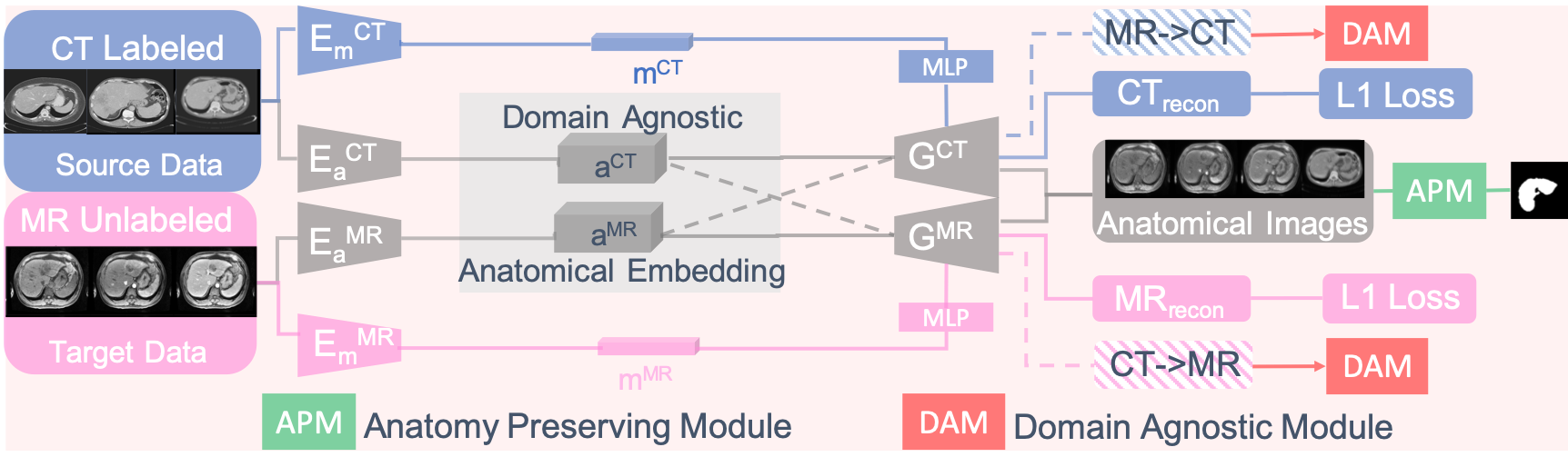}
\end{center}
   \caption{(Best viewed in color) The end-to-end DALACE pipeline to learn domain-agnostic anatomical embeddings.  The solid line shows the self-reconstruction process while the dotted line shows the cross-reconstruction/modality-transfer generation process.}
\label{fig:pipeline}
\end{figure*}
\subsection{End-to-End Pipeline}
Fig. \ref{fig:pipeline} shows the end-to-end DALACE pipeline to learn a domain-agnostic anatomical embedding, which is invariant to domains but discriminative of the classes for the segmentation task. Input CT and MR images are denoted as $X_{CT}$ and $X_{MR}$. Inspired by the MUNIT \cite{huang2018multimodal} model and DADR \cite{yang2019domain} model, DALACE consists of two anatomy encoders $E_a^{CT}$ and $E_a^{MR}$, two modality encoders $E_m^{MR}$ and $E_m^{CT}$, and two style-based generators with multi-layer perceptron (MLP) and adaptive instance normalization (AdaIN) \cite{karras2019style} $G^{CT}$ and $G^{MR}$. We propose the Anatomy Preserving Module (APM) and Domain Agnostic Module (DAM) to generate domain-agnostic anatomical images for the DA and DAL tasks.

To start the pipeline, both source data $x_{CT}$ and target data $x_{MR}$ are fed into the encoders ($E_a^{CT}$, $E_m^{CT}$, $E_a^{MR}$ and $E_m^{MR}$) and embedded into anatomy codes $a^{CT}$ and $a^{MR}$ (feature maps) and modality codes $m^{CT}$ and $m^{MR}$ (vectors). In the next step, anatomy codes and modality codes are fed into style-based generators $G^{CT}$ and $G^{MR}$ for self-reconstruction via optimizing the $L_{img}$ term in equation \eqref{eq:recon}.  Then modality codes $m^{CT}$ and $m^{MR}$ are swapped and together with the original anatomy codes are fed into style-based generators for cross-reconstruction/modality-transfer generation, which is contrained by the $L_{latent}$ loss term in equation \eqref{eq:recon}. Please refer to \eqref{eq:img} and \eqref{eq:latent} for details about $L_{img}$ and $L_{latent}$. Expectation is taken with respect to $x_{CT} \sim X_{CT}$ and $x_{MR} \sim X_{MR}$.


\begin{equation}\label{eq:recon}
\begin{split}
L_{recon} &= \alpha L_{img} + \beta L_{latent}
\\&= \alpha (L_{CT}+L_{MR}) \\&+ \beta (L_{a}^{CT}+L_{m}^{CT}+L_{a}^{MR}+L_{m}^{MR})
\end{split}
\end{equation}


\begin{equation}\label{eq:img}
\begin{split}
&\;\;\;\;\,L_{CT}+L_{MR}
\\&=\mathbb{E}||G^{CT}(E_{a}^{CT}(x_{CT}),E_m^{CT}(x_{CT}))-x_{CT}||_1 \\&+ \mathbb{E}||G^{MR}(E_{a}^{MR}(x_{MR}),E_m^{MR}(x_{MR}))-x_{MR}||_1
\end{split}
\end{equation}


\begin{equation}\label{eq:latent}
\begin{split}
&\;\;\;\;\,L_{a}^{CT}+L_{m}^{CT}+L_{a}^{MR}+L_{m}^{MR}
\\&=||E_{a}^{MR}(x_{CT\rightarrow MR})-a^{CT}||_1
\\&+||E_{m}^{CT}(x_{MR\rightarrow CT})-m^{CT}||_1
\\&+||E_{m}^{MR}(x_{CT\rightarrow MR})-m^{MR}||_1
\\&+||E_{a}^{CT}(x_{MR\rightarrow CT})-a^{MR}||_1 
\end{split}
\end{equation}

To generate anatomy-preserving domain-agnostic images, only anatomy codes alone are fed into the generators without modality codes. DAM encourages the anatomy embedding to be domain-agnostic by adversarial training while APM encourages the anatomy embedding to be anatomy-preserving by adversarial training \cite{huang2018multimodal, zhang2018task}. In this way, the model is designed to learn meaningful and interpretable disentangled representations, thus helping us to understand the learned representations and the model better.  

\subsection{Feedback Supervision Modules}
\subsubsection{Domain Agnostic Module}

This module encourages the embedding to be domain-agnostic in an adversarial training way. It consists of two discriminators $D^{CT}$ and $D^{MR}$, which try to discriminate between real CT $X_{CT}$ and fake CT transferred from MR $X_{MR\rightarrow CT}$ and real MR $X_{MR}$ and fake MR transferred from CT $X_{CT\rightarrow MR}$, respectively. The discriminators compete with encoders and style-based generators to encourage the disentanglement of modality and anatomical information by driving modality information into the modality codes, thus forcing the anatomy embedding to be domain-agnostic. Please see Equation \eqref{eq:DAM} \eqref{eq:CTtoMR} \eqref{eq:MRtoCT} for details.

\begin{figure}
\begin{center}
\includegraphics[width=0.525\linewidth]{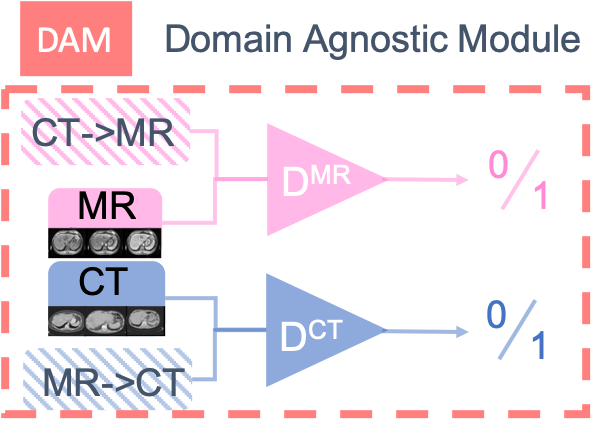}
\end{center}
   \caption{Domain-Agnostic Module, which encourages the embedding to be domain-agnostic by adversarial training.}
\label{fig:long}
\label{fig:onecol}
\end{figure}

\begin{equation}\label{eq:DAM}
L_{adv}^{cross}=L_{adv}^{CT\rightarrow MR}+L_{adv}^{MR\rightarrow CT}
\end{equation}
\begin{equation}\label{eq:CTtoMR}
\begin{split}
L_{adv}^{CT\rightarrow MR}
&= \mathbb{E}[log(1-D^{MR}(x_{CT\rightarrow MR}))]\\&+ \mathbb{E}[log(D^{MR}(x_{MR}))]
\end{split}
\end{equation}
\begin{equation}\label{eq:MRtoCT}
\begin{split}
L_{adv}^{MR\rightarrow CT}
&=\mathbb{E}[log(1-D^{CT}(x_{MR\rightarrow CT}))]\\&+\mathbb{E}[log(D^{CT}(x_{CT}))]
\end{split}
\end{equation}

\subsubsection{Anatomy Preserving Module (APM)}
The Anatomy Preserving Module helps the embedding to preserve and align high-level semantic information for different modalities. It consists of two steps. In the first step, both anatomical images from CT and MR, $X_{a}^{CT}$ and $X_{a}^{MR}$, are fed into a segmentation module $S$, i.e. a U-Net based model, to generate segmentation masks $\hat{M}_{a}^{CT}$ and $\hat{M}_{a}^{MR}$ for both $X_{a}^{CT}$ and $X_{a}^{MR}$. For $\hat{M}_{a}^{CT}$, we compute the pixel-wise cross entropy loss (Equation \eqref{eq:CE}) between $\hat{M}_{a}^{CT}$ and the ground truth mask of original CT image $M_{a}^{CT}$ to encourage the encoders and style-based generators to keep the anatomy information. For $\hat{M}_{a}^{MR}$, we train a conditional GAN to differentiate between the pair of $X_{a}^{MR}$ and $\hat{M}_{a}^{MR}$ and the pair of $X_{a}^{CT}$ and $\hat{M}_{a}^{CT}$ (Equation \eqref{eq:CGAN}), thus encouraging the pair of anatomical images and prediction masks originally from CT and MR to be nondifferentiable so that the anatomical images from MR will be anatomy preserving in an adversarial way.

\begin{figure}
\begin{center}
\includegraphics[width=0.725\linewidth]{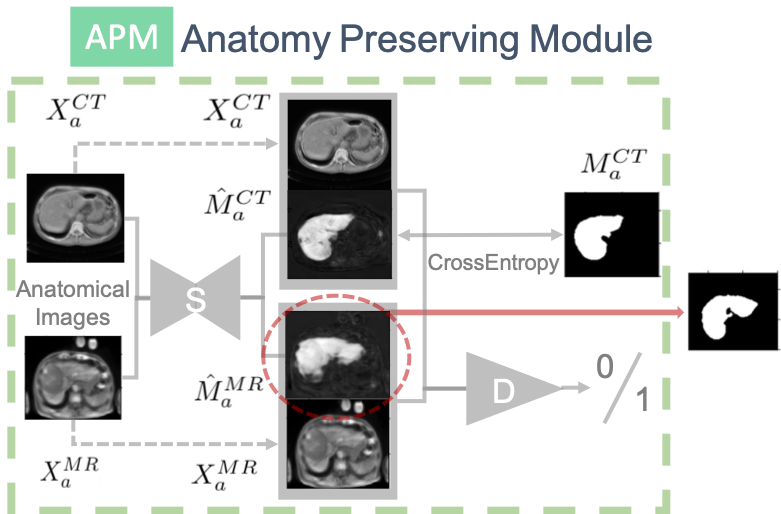}
\end{center}
   \caption{Anatomy-Preserving Module, which encourages the embedding to be anatomy-preserving by adversarial training. D is the discriminator, S is the U-Net segmentation module.}
\label{fig:long}
\label{fig:onecol}
\end{figure}

\begin{equation}\label{eq:CE}
L_{CE} = -\Sigma \; y_{true}log(y_{pred})
\end{equation}

\begin{equation}\label{eq:CGAN}
\begin{split}
L_{adv}^{pair}
&= \mathbb{E}[log(1-D(x_a^{MR}, \hat{M}_a^{MR}))]\\&+ \mathbb{E}[log(D(x_a^{CT}, \hat{M}_a^{CT}))]
\end{split}
\end{equation}


\subsection{Implementations Details}
Anatomy encoders consist of 1 convolutional layer of stride 1 with 64 filters, 2 convolutional layers of stride 2 with 128, 256 filters respectively and 4 residual layers with 256 filters followed by batch normalization, while modality encoders are composed of 1 convolutional layer of stride 1 with 64 filters, 4 convolutional layers of stride 2 with 128, 256, 256, and 256 filters, a global average pooling layer, and a fully-connected layer with 8 filters without any batch normalization. Style-based generators with MLP take the anatomy codes (feature maps of size 64x64x256) and modality codes (vector of length 8) as inputs, which consist of 4 residual layers with 256 filters, 2 upsampling layers of 2x, and 1 convolutional layer of stride 1. The modality codes are used as inputs to the MLP to generate affine transformation parameters. Residual blocks in the style-based generators are equipped with an Adaptive Instance Normalization (AdaIN) layer to take the affine transformation parameters from the modality codes via the MLP. Discriminators are convolutional neural networks for binary classification. As for the DAM and APM modules, the segmentation network is a standard U-Net \cite{ronneberger2015u} architecture and the discriminators are also convolutional binary classifiers. 
The Adam optimizer \cite{kingma2014adam} is used for optimization. To update the parameters in the DALACE model, 
First, $alpha$ and $\beta$ are set as 2.5 and 0.01 for minimization of the loss function in equation \eqref{eq:recon}: $\min_{E_a, E_m, G}\, L_{recon}= \alpha L_{img} + \beta L_{latent}$.
Second, adversarial training is applied for the loss function in equation \eqref{eq:DAM}: $\min_{E_a, E_m, G} \, \max_{D} \, L_{adv}^{cross}$.
Thhird, loss functions in equation \eqref{eq:CE} and \eqref{eq:CGAN} are optimized as $\min_S\,L_{CE}$, $\min_{E_a, G} \, \max_{D} \, L_{adv}^{pair}$, where $S$ denotes the segmentation module in APM.  
  
Learning rate is set as 0.001 except 0.0001 for $\min_{E_a, G} \, \max_{D} \, L_{adv}^{pair}$. In total, 2600 epochs are trained for each fold. In the first 600 epochs, $L_{adv}^{pair}$ is not optimized. Experiments were conducted on two Nvidia 1080ti GPUs. The training time each fold is $\sim2.5\,h$. The testing time each fold is within a minute.

\section{Experimental Results}
\subsection{Data and Preprocessing}
\begin{figure}
\begin{center}
\includegraphics[width=0.5\linewidth]{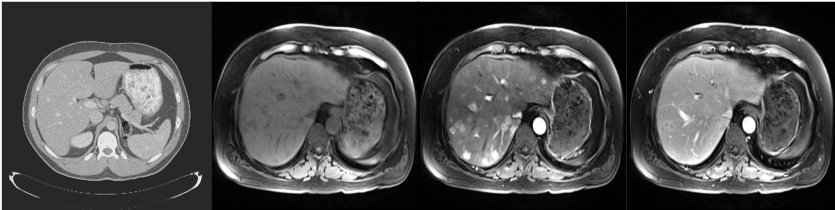}
\end{center}
   \caption{Examples of images from different modalities, from left to right: CT, pre-contrast phase MR, 20s post-contrast phase MR and 70s post-contrast phase MR.}
\label{fig:DATA}
\end{figure}

We tested our DALACE model on slices from unpaired CT and MR scans: 130 CT scans from the LiTS challenge at ISBI and MICCAI 2017 \cite{christ2017lits} and multi-phasic MR scans from 20 patients at a local medical center, including pre-contrast phase MR, 20s post-contrast phase (arterial phase) MR and 70s post-contrast phase (venous phase) MR. Please see Fig. \ref{fig:DATA} for image examples. Not only is the huge domain shift from CT to MR observed, the domain shifts between multi-phasic MR images can not be neglected. The multi-phasic MR dataset of 20 patients was collected with Institutional Review Board (IRB) approval and manual liver segmentation masks were created by a radiology expert. Both MR and CT data are normalized and resliced to be isotropic in three dimensions. Bias field correction is applied on MR data. For all the experiments, both CT and MR datasets are partitioned into 5-folds for cross-validation purposes.




\subsection{Domain Adaptation}
For the DA task, to transfer knowledge from CT to pre-contrast phase MR, competing models are trained with labeled CT images and unlabeled pre-contrast phase MR images. Model performance was assessed using the dice similarity coefficient (DSC) between true and predicted liver segmentations.

To have a better sense of understanding of the data and the DA task, we have a supervised U-Net trained and tested on the small pre-contrast phase MR dataset to serve as the upperbound. Another supervised U-Net is trained on CT and tested on pre-phase MR to serve as the lowerbound for each task. Please see Table \ref{tab:Seg_DAL} for details. The upperbound might be lower than the actual upperbound since the training MR data for 5-fold cross-validation is small and noisy. Compared to the MR data, CT data is much more available and robust to artifacts.

\textbf{\textit{Settings}} For each cross-validation split, four folds of CT data with segmentation masks and pre-contrast MR data without segmentation masks are used to train, and one fold of pre-contrast MR data without segmentation masks is used to test. The state-of-the-art models CycleGAN \cite{zhu2017unpaired}, TD-GAN \cite{zhang2018task}, and DADR \cite{yang2019domain} are trained with the same partition of data for the DA task. DALACE finds a shared space to embed both CT and MR and transfers both modalities into anatomical images while CycleGAN and TD-GAN tries to transfer directly between CT and MR.

\begin{figure}
\begin{center}
\includegraphics[width=0.775\linewidth]{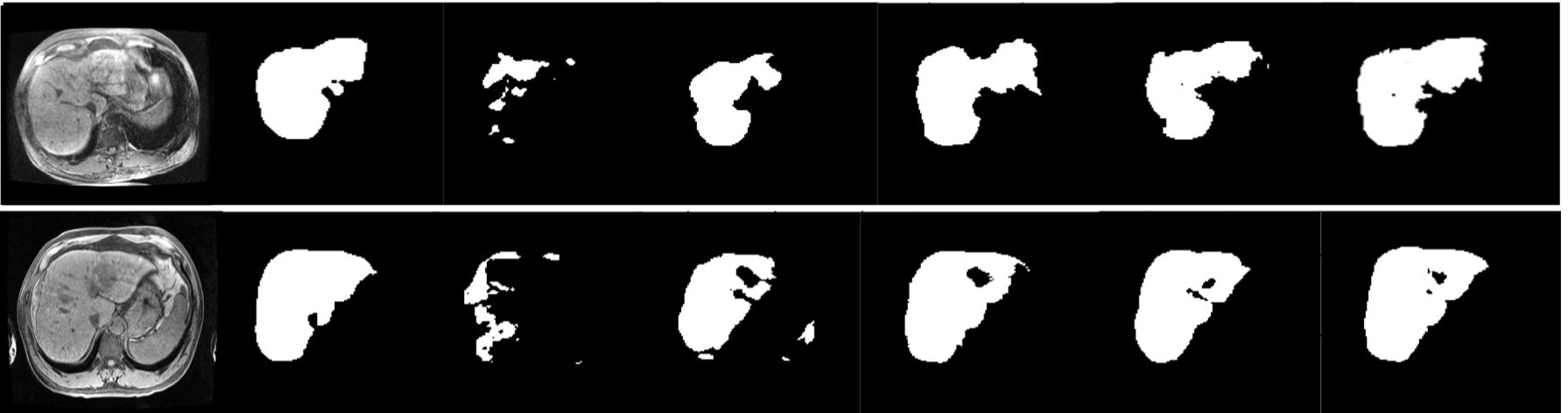}
\end{center}
   \caption{Two examples of DA task for cross-modality liver segmentation with different methods. From left to right: original pre-contrast phase MR images, ground truth masks, U-Net w/o DA results, CycleGAN results, TD-GAN results, DADR results, DALACE results.}
\label{fig:Seg}
\end{figure}

\textbf{\textit{Results}} As shown in Table \ref{tab:Seg_DA}, our DALACE model outperforms the current state-of-the-art models with DSC of 0.847 compared to DSC of 0.721 for CycleGAN, 0.793 for TD-GAN, and 0.806 for DADR. Please see Fig. \ref{fig:Seg} for visual comparison of qualitative results from different models on cross-modality liver segmentation with DA. 

\begin{table}
\begin{center}
\scalebox{0.725}{
\begin{tabular}{|l|l|}
\hline
DA task & DSC (std) \\
\hline
lowerbound & 0.260 (0.072)\\
upperbound & 0.869 (0.044)\\
\hline
\end{tabular}
\quad
\begin{tabular}{|l|l|}
\hline
Method & DSC (std) \\
\hline
CycleGAN \cite{zhu2017unpaired} & 0.721 (0.049)\\
TD-GAN \cite{zhang2018task}& 0.793 (0.066)\\
DADR \cite{yang2019domain}& 0.806 (0.035)\\
\textbf{DALACE} & \textbf{0.847 (0.041)}\\

\hline
\end{tabular}}
\end{center}
\caption{DA results. Estimated lowerbound and upperbound for cross-modality liver segmentation with DA. Comparison of
segmentation results for domain adaptation with different models. Our DALACE outperforms other methods.}
\label{tab:Seg_DA}
\end{table}

\subsection{Domain Agnostic Learning}
For the DAL task, to transfer knowledge from CT to multi-phasic MR, competing models are trained with labeled CT images and unlabeled MR images in three different phases. 

To better assess performance on the DAL task, we have a supervised U-Net trained and tested on MR from each phase separately to serve as the upperbound. Another supervised U-Net is trained on CT and tested on MR from all phases to serve as the lowerbound. Please see Table \ref{tab:Seg_DAL} for details. The upperbound might be lower than the actual upperbound given the noisy and small MR dataset for training with 5-fold cross-validation. Among different phases, liver in the arterial phase MR is more visually inhomogeneous than liver in other MR phases, which might lead to downgraded performance.

\textbf{\textit{Settings}} For each cross-validation split, four folds of CT data with segmentation masks and multi-phasic MR data including pre-contrast phase, 20s post-contrast phase and 70s post-contrast phase without segmentation masks are used to train, and one fold of the multi-phasic MR data is used to test. The state-of-the-art models CycleGAN \cite{zhu2017unpaired}, TD-GAN \cite{zhang2018task} and DADR \cite{yang2019domain} are trained with the same partition of data for the DAL task. 

\textbf{\textit{Results}} As shown in Table \ref{tab:Seg_DAL}, our DALACE model outperforms the current state-of-art models with DSC of 0.794 compared to DSC of 0.522 for CycleGAN, 0.719 for TD-GAN, and 0.742 for DADR. The shared embedding space from our DALACE model is modality-invariant to CT and multi-phasic MR, thus it is effective on the DAL task where target data is from heterogenous mixed domains. CycleGAN and TD-GAN performed badly in transferring between CT and multi-phasic MR since they are assuming multi-phasic MR to be from the same domain. DADR assumes mixed domains, but does not enforce anatomy-consistent representations, which results in lower performance compared to our DALACE model.

\begin{table}
\begin{center}
\scalebox{0.725}{
\begin{tabular}{|l|l|}
\hline
DAL task & DSC (std) \\
\hline
lowerbound & 0.228 (0.130)\\
upperbound &  0.823 (0.057)\\
\hline
\end{tabular}
\quad
\begin{tabular}{|l|l|}
\hline
Method & DSC (std) \\
\hline
CycleGAN \cite{zhu2017unpaired} & 0.522 (0.064)\\
TD-GAN \cite{zhang2018task}& 0.719 (0.089)\\
DADR \cite{yang2019domain}&  0.742 (0.045) \\
\textbf{DALACE} & \textbf{0.794 (0.044)}\\
\hline
\end{tabular}}
\end{center}
\caption{DAL results. Estimated lowerbound and upperbound for cross-modality liver segmentation with DAL. Comparison of
segmentation results for domain adaptation with different models. Our DALACE generalizes well to the DAL task compared to other methods.}
\label{tab:Seg_DAL}
\end{table}

\subsection{Joint Learning}
For joint learning, instead of transferring knowledge from CT to MR, knowledge from CT and MR are jointly learned to get a better model on both CT and MR. Specifically, not only do CT images have ground truth masks, but also MR images have ground truth masks for training. 


\textbf{\textit{Settings}} CT and pre-contrast phase MR are used in this experiment. For each cross-validation split, four folds of CT with segmentation masks and four folds of MR with segmentation masks are used to train the DALACE model, and the other one fold of CT and MR is used to test the model. U-Net trained on four folds of CT with segmentation masks and tested on the other one fold of CT and U-Net trained on four folds of MR with segmentation masks and tested on the other one fold of MR were used for comparison. 

\textbf{\textit{Results}}
As shown in Table \ref{tab:Joint}, the DALACE model for joint learning simultaneously outperforms the fully-supervised U-Net models separately trained and tested on each modality, with DSC of 0.911 tested on CT and 0.907 tested on MR compared to 0.901 on CT and 0.869 on MR using fully-supervised U-Net. Of note, 0.869 (0.044) is the estimated upperbound for DA tasks from Table \ref{tab:Seg_DA}. Overall, our DALACE model outperformed other methods for the joint learning task, especially in terms of MR tested DSC, which is of most interest. Only two methods DADR and DALACE in joint learning exceeded the upperbound for DA and showed synergy from effectively intergrating information from both CT and MR. Since our DALACE model for joint learning uses the domain-agnostic images of CT and MR as the inputs for the segmentation module, it shows that the DALACE model successfully disentangles the anatomy information from modality information. To achieve the task of liver segmentation, it does not necessarily require information from modality codes, but only anatomical information is relevant to the segmentation task.

\begin{table}
\begin{center}
\scalebox{0.725}{
\begin{tabular}{|l|l|l|}
\hline
Method & CT tested DSC  & MR tested DSC \\
\hline\hline
CT trained U-Net & 0.901 (0.020) & 0.260 (0.072)\\
MR  trained U-Net & 0.134 (0.091) & 0.869 (0.044)\\
CT\&MR trained U-Net & 0.835 (0.035)& 0.590 (0.098)\\
Joint CT\&MR CycleGAN & 0.870 (0.023) & 0.846 (0.048)\\ 
Joint CT\&MR TD-GAN & 0.880 (0.018)& 0.863 (0.029)\\ 
Joint CT\&MR DADR & \textbf{0.912 (0.012)}& 0.891 (0.040) \\ 
\textbf{Joint CT\&MR DALACE} & 0.911 (0.013) & \textbf{0.907 (0.049)}\\

\hline
\end{tabular}}
\end{center}
\caption{Joint learning results. Results of joint learning models and comparison with fully-supervised U-Net models on each modality.}
\label{tab:Joint}
\end{table}

\section{Analysis}
\subsection{Results Analysis}
We tested our DALACE model on unpaired CT and MR data in three experiments and showed that DALACE is superior to the current state-of-the-art models in the literature such as CycleGAN, TD-GAN and DADR. The DALACE model, which works on both domain-transfer and task-transfer to learn a disentangled representation, not only aims to be invariant to different modalities but also preserves anatomical structures. In the DAL experiment, the main challenge is that target data come from multiple target domains, which violates the assumptions made by CycleGAN and TD-GAN. Features in CycleGAN and TD-GAN are highly entangled so that it is hard to learn a domain-invariant representation given multiple domains. However, the DALACE model generalizes easily to the DAL experimental settings and demonstrates superior performance in terms of DSC score due to disentanglement learning. Compared to DADR, the DALACE model achieved improved performance for the downstream tasks through explicitly enforcing semantic consistency. In the Joint-Learning experiment, we show the potential of our DALACE model to integrate different modalities, which shows the meaningful disentangled representations from each domain are domain-agnostic and aligned to preserve the anatomy structures.

\subsection{Visualization of Disentanglement}
Through the above experiments, we have shown that our DALACE model outperforms the current state-of-the-art models on both DA and DAL tasks. In this section, we will show that anatomy information and modality information are disentangled by the DALACE model through visualization of domain-agnostic images and modality-transferred images.

\textit{\textbf{Domain-Agnostic Images}}
\begin{figure}
\begin{center}
\includegraphics[width=0.9\linewidth]{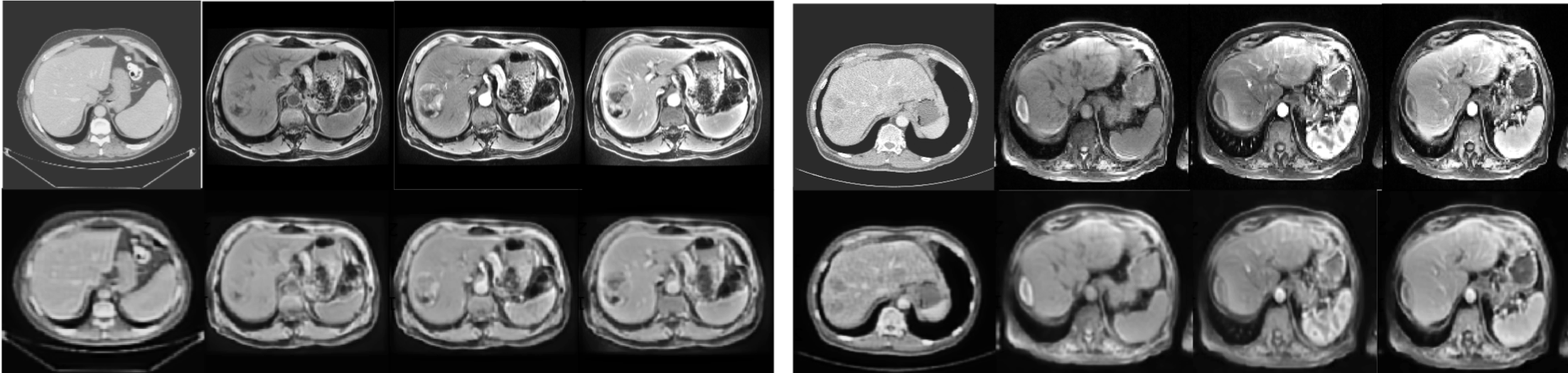}
\end{center}
   \caption{Two sets of examples of domain-agnostic images. In each set, the first row from right to left is CT, pre-contrast MR, 20s post-contrast MR, and 70s post-contrast MR, and the second row is its corresponding domain-agnostic images.}
\label{fig:DAIMG}
\end{figure}
To generate domain-agnostic images, CT and MR images are embedded by encoders into anatomy codes and modality codes.  Then only anatomy codes are fed into style-based generators without modality codes to get the outputs as domain-agnostic images. Please see Fig. \ref{fig:DAIMG} for domain-agnostic anatomical images generated by anatomy codes from different domains including CT and multi-phasic MR. As demonstrated in the figure, the modality information is erased while the anatomical structures are preserved in the domain-agnostic images. In other words, the anatomy information is extracted and preserved in the domain-agnostic anatomical embeddings. 

\begin{figure}
\begin{center}
\includegraphics[width=0.36\linewidth]{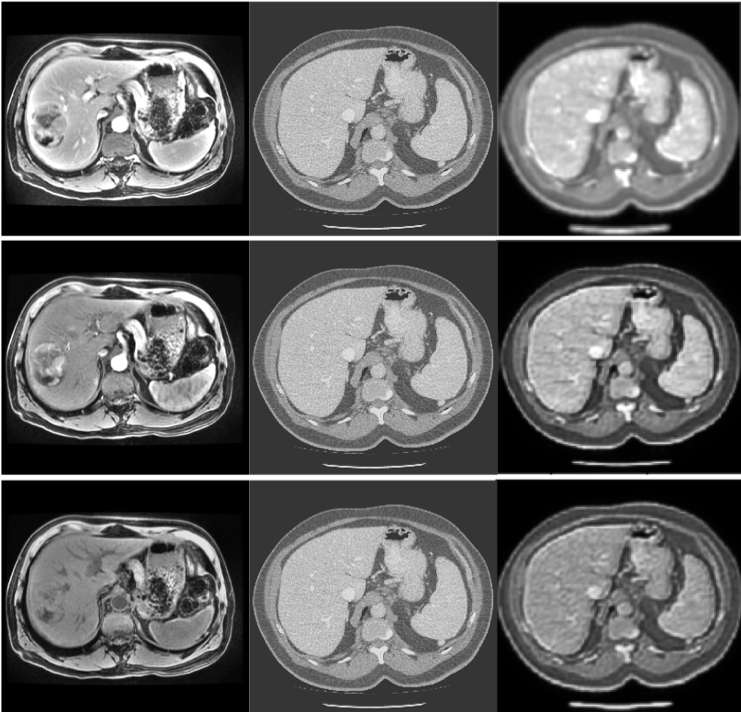}
\end{center}
   \caption{CT images are transferred to multi-phasic MR images in three phases. From left to right, each column is the multi-phasic MR images (from top to bottom: 70s post-contrast phase MR, 20s post-contrast phase MR, pre-contrast phase MR), the CT images, the modality-transferred images with anatomy structure from the CT images in the second column and modality rendering from the multi-phasic MR images in the first column.}
\label{fig:STIMG}
\end{figure}

\textit{\textbf{Modality-Transferred Images}}
To perform modality transfer, both input images and reference images are embedded by encoders into anatomy codes and modality codes. Then we maintain the anatomy codes from input images and modality codes from reference images and feed them into the style-based generators to get modality-transferred images. The generated modality-transferred images will inherit the anatomy structure from input images and modality rendering from reference images. Please see Fig. \ref{fig:STIMG} for CT images transferred to multi-phasic MR images in three phases. It shows the successful disentanglement of modality information into modality code.

\subsection{Interpretation}
According to the categories for deep learning model explanation methods in \cite{gilpin2018explaining}, the DALACE model is designed to be easier to interpret by explicitly learning meaningful and interpretable representations. 
The successful disentanglement of anatomy and modality information, as shown in Fig. \ref{fig:DAIMG} and Fig. \ref{fig:STIMG}, adds transparency to the black-box model. Furthermore, in the previous experiments, it was shown that the learned meaningful and interpretable representation is able to generalize and is useful for reconstruction and downstream tasks.  
\section{Ablation Studies}
Recent work on disentanglement learning \cite{locatello2019challenging} suggests that, besides demonstrating the successful disentanglement, two important directions of future research are: (1) to investigate the concrete benefits of enforcing disentanglement learning for downstream tasks. (2) to explicitly discuss the role of supervision on disentanglement. Ablation studies are performed on our model to analyze the role of the components in our proposed model, in accordance with the above two points.

\subsection{Effectiveness of Disentanglement}
To investigate the concrete benefits of enforcing disentanglement of the learned representations, we took out the disentanglement from our model by replacing the anatomy encoders, modality encoders and style-based generators with CycleGAN and the other parts of the model remain the same, except that there will be no domain-agnostic images and direct modality-transfer is applied between CT and MR, which is essentially the TD-GAN \cite{zhang2018task} model with the segment module pretrained on CT. The ablation experiment showed that, without the disentanglement component, the performance decreased from 0.847 to 0.793 for the DA task and from 0.794 to 0.719 for the DAL task, which indicates that the disentanglement benefits the performance of downstream tasks. 


\subsection{Role of Supervision on Disentanglement}
\begin{table}
\begin{center}
\scalebox{0.725}{
\begin{tabular}{|l|l|l|l|}
\hline
APM & DAM & DSC for DA & DSC for DAL \\
\hline\hline
 & \checkmark & 0.806 (0.035) &  0.742 (0.041) \\
 \hline
 \checkmark &  & 0.776 (0.078) &  0.702 (0.132) \\
 \hline
\textbf{\checkmark} &\textbf{\checkmark} & \textbf{0.847 (0.041)}&\textbf{0.794 (0.044)}\\
\hline
\end{tabular}}
\end{center}
\caption{Ablation studies on the Role of Supervision on Disentanglement of Anatomy Preserving Module (APM) and Domain Agnostic Module (DAM).}
\label{tab:supervision}
\end{table}
To be explicit about the role of supervision for disentanglement, as well as to investigate the role of APM and DAM in the DALACE model, we take out the APM and DAM part respectively. Taking out APM will separate the end-to-end DALACE model into a two-stage model without enforcing semantic consistency, which is essentially DADR \cite{yang2019domain}. Taking out DAM will result in weakening the model's ability to learn a domain agnostic representation, thus degrading the performance. Please see Table \ref{tab:supervision} for details. It shows the important role of supervision on disentanglement and performance.  

\section{Conclusions and Limitations}
For medical image analysis, in practice, it is expensive and time consuming to collect and annotate medical images. DA can be an effective solution for generalization of deep learning models for medical image analysis. However, target data itself can come from different scanners, medical sites, protocols and modalities with domain shifts, demonstrating the importance of the proposed DAL task. In addition, each modality plays a unique role in the diagnosis and after-treatment follow-up. An accurate model for the DAL task not only solves the problem of scarcity of labeled training data for medical image analysis using deep learning, but also it will improve the current clinical workflow and greatly help the integration of different modalities. 

This work explicitly proposed the DAL task for medical image analysis and introduced DALACE, an end-to-end trainable model which utilizes disentanglement to preserve the anatomical information and promote domain adaptation to the new DAL task. Through ablation studies, we explicitly investigated the effectiveness of disentanglement and the role of supervision for disentangled representation that is domain agnostic  and anatomy preserving.  By visualization, we showed that the disentanglement promotes the interpretability of the learned representation. 

While DALACE is proposed to tackle the DA and DAL tasks, it also has the potential to realize style transfer. Getting one model that works on style transfer, DA and DAL tasks is difficult but
 desirable and an interesting direction for future work. Without paired CT and MR to serve as ground truth, style transfer results are difficult to be quantitatively evaluated. The joint learning experiment also points to a potential direction for future studies, the integration of modalities.

{\small
\bibliographystyle{ieee}
\bibliography{egbib}
}

\end{document}